\documentclass[11pt, a4paper]{article}

\usepackage{geometry}
\usepackage{amsmath}
\usepackage{amsfonts}
\usepackage{amssymb}
\usepackage{physics}
\usepackage{tensor}
\usepackage{mathtools}
\usepackage{mathrsfs}
\usepackage{graphicx}
\graphicspath{{.}}
\DeclareGraphicsExtensions{.pdf}

\usepackage{hyperref}
\hypersetup{ 
	colorlinks=true,
	linkcolor=blue,
}

\begin{document}
\title{Electrostatics and Riemann Surfaces}
\author{Spencer Tamagni\footnote{stamagni8@knights.ucf.edu} ~and~ Costas Efthimiou\footnote{costas@physics.ucf.edu}\\
\small Department of Physics, University of Central Florida, Orlando, FL 32816}
\date{May 31, 2020}

\maketitle	

\begin{abstract}
Using techniques from geometry and complex analysis in their simplest form, we present a derivation of electric fields on  surfaces with non-trivial topology. A byproduct of this analysis is an intuitive visualization of elliptic functions when their argument is complex-valued. The underlying connections between these techniques and the theory of Riemann surfaces are also explained. Our goal is to provide students and instructors a quick reference article for an extraordinary topic that is not included in the standard books.
 \end{abstract}

\section{Introduction}
Concepts from topology and geometry are ubiquitous in physics. In the era of the topological insulator, this is quite evident. Moreover, geometrical and physical intuition are intimately linked: each aids in elucidating the key concepts of the other. 

Unfortunately, modern geometry is somewhat dense, and is inaccessible at the undergraduate level. In spite of this, the underlying ideas are concrete and intuitive. In addition, geometrical concepts are usually introduced to physicists in relatively sophisticated physical contexts, typically involving general relativity or quantum field theory. We aim to (partially) remedy this issue by considering a 19th century physics problem which illustrates geometrical concepts --- the formulation and solution of electrostatics on compact\footnote{The exact mathematical definition of a compact surface is not necessary for our purposes. The reader may imagine such a surface as a finite surface (i.e. one that does not go to infinity) and without boundary (i.e. without a borderline that belongs to it).} surfaces. In particular, we focus on the simplest cases of the sphere and torus.

We believe that working out this problem in detail is of pedagogical value because electrostatics is a standard subject which students of physics are held accountable for knowing well. Therefore, we may use it as a laboratory for exploring more unfamiliar geometrical ideas, which turn out to involve the theory of Riemann surfaces. This also establishes that geometrical concepts appear even at the introductory level in physics, although we tend to overlook them. 

The main technical tools we employ are the notions of conformal mapping and Riemann surfaces, each of which we explain in a self-contained way. Of course, neither of these techniques are new, and they appear in various contexts throughout the literature. See, for example, \cite{levin}, \cite{staunton}, \cite{weigel} for applications of conformal transformations in the more standard context of boundary value problems in the plane and \cite{MittagStephen} for  some applications in classical and quantum mechanics. 
Riemann surfaces are ubiquitous in mathematical physics; for interesting applications to some spectral problems (which one may think of in some sense as a ``complexified quantum mechanics"), see \cite{burton}, \cite{sokolovski}.  Also, the relation of electromagnetism and fluid mechanics has a long history and has been exploited
to construct electric fields from flows or the other way around. In particular, in \cite{EsparzaEtAl}, the authors construct 2-dimensional vortices on a sphere using the vector velocity potential.
Our main contribution is an explanation of how to use these techniques to solve for electric fields on domains of nontrivial topology. We show how simple, albeit clever, geometric constructions may be used to drastically simplify what would be quite difficult analytic problems. This allows us to obtain closed form expressions for electrostatic potentials in terms of the complex-analytic data associated to a Riemann surface.

We have structured this paper as follows: in the first section, we provide a brief review of some background mathematics from complex analysis. In section 2, we explore some topological subtleties of electrostatics on compact spaces. In section 3, we explain how to use the stereographic projection to obtain electric fields on the sphere from familiar fields on the plane. Section 4 concerns fields on the torus, where we use geometry to explain a beautiful connection between electrostatics and the theory of elliptic functions. 

In order to remain concise, we avoid exhaustive and general mathematical constructions, choosing instead to develop all concepts by example. In the concluding section, we direct the interested reader to the relevant literature to learn the general formulation of the concepts we illustrate.

\section{Complex Variables}
Our most essential tool will be the concept of a holomorphic function. One typically encounters such functions in a course on Mathematical Methods in physics and, of course, on Complex Analysis in mathematics. We will review the fundamental results that we will need in this section.

If we introduce the complex coordinate $z = x + iy$ on the plane, we have that 
\begin{align*}
x = \frac{z + \overline{z}}{2}, \quad
y = \frac{z - \overline{z}}{2i},
\end{align*}
where $\overline{z} = x -iy$ is the complex conjugate. Due to these formulas, a general function $f(x, y)$ will depend on both $z$ and $\overline{z}$ when expressed in complex coordinates, $f(z,\overline z)$.

The functions for which $f(z, \overline{z})$ is in fact independent of $\overline{z}$ are known as \textit{holomorphic}, written as $f(z)$, and satisfy the trivial Cauchy-Riemann condition
\begin{equation} 
\label{1}
\pdv{f}{\overline{z}} = 0.
\end{equation}
We may write $f = u + iv$ and then we have 
\begin{subequations}
\label{eq:map}
\begin{eqnarray}
 u & = & u(x, y); \\
 v & = & v(x, y).
\end{eqnarray}
\end{subequations}
Taking the real and imaginary parts of \eqref{1} gives the more conventional statement of the Cauchy-Riemann conditions,
\begin{subequations}
\begin{eqnarray*}
 {\partial u\over\partial x}  & = & +{\partial v\over\partial y} , \\
 {\partial u\over\partial y}  & = & -{\partial v\over\partial x} ,
\end{eqnarray*}
\end{subequations}
a simple exercise we encourage for the reader.

One useful property of holomorphic functions is the following: regarded as a transformation, equations \eqref{eq:map}
give a mapping from the $xy$-plane to the $uv$-plane. This mapping is \textit{conformal}, or angle-preserving. Precisely, if two curves $C$ and $C'$ in the $xy$-plane meet at some point $P$ with angle $\theta$, the corresponding curves $f(C)$, $f(C')$ meet at $f(P)$ in the $uv$-plane \textit{with the same angle $\theta$}. 

Another useful property of holomorphic functions is that the families of curves $u(x, y) = \text{const.}$ and $v(x, y) = \text{const.}$ are orthogonal, that is, the curves always meet at right angles. Again, the reader is encouraged to prove this, using the Cauchy-Riemann conditions.

A useful tool for both analysis and visualization of two-dimensional fields is the so-called ``complex potential", which we proceed to define and explain.

The main equation for electrostatics is Gauss' law. In differential form and for  points that have no charge, the law is  $\div \vec{E} = 0$. We traditionally introduce the scalar potential $\varphi$ by $\vec{E} = - \grad \varphi$. By direct substitution in Gauss' law, this results in  Laplace's equation 
\begin{equation*}
        \laplacian \varphi = 0. 
\end{equation*}
In two dimensions, or in a three dimensional situation with translation symmetry in one direction, using the complex coordinate $z = x +i y$, the Laplacian takes the form
\begin{equation*}
       \laplacian = \frac{\partial^2}{\partial x^2} + \frac{\partial^2}{\partial y^2} = \Big( \pdv{}{x} - i \pdv{}{y} \Big) \Big( \pdv{}{x} + i \pdv{}{y} \Big) 
                       = 4 \partial_z \partial_{\overline{z}}.
\end{equation*}
This means that Laplace's equation becomes 
$$
       \partial_z \partial_{\overline{z}} \varphi = 0,
$$
with the general solution $  \varphi(z, \overline{z}) =  F(z) + \overline{G(z)}$, where $F(z), G(z)$ are any arbitrary holomorphic functions\footnote{This ensures that $\overline{G(z)}$ is antiholomorphic, that is, only depends on $\overline{z}$.}.
 Since the potential must be real,
\begin{equation*}
    \varphi(z, \overline{z}) = \frac{f(z)+\overline{f(z)}}{2} .
\end{equation*} 
In other words, $\varphi$ is the real part of the holomorphic function $f(z)$ which is known as the \textit{complex potential}. It will be most useful for us to deal with $f$ directly instead of $\varphi$. 

Note that the equipotential surfaces are simply the solutions to $\Re f(z) = \text{const.}$ The electric field is of course normal to the equipotential surfaces.  
It then follows that  the solutions to $\Im f(z) = \text{const.}$ --- which form a family of curves orthogonal to the equipotential surfaces --- are the electric field lines.
Hence, knowing the complex potential $f(z)$ allows one to reconstruct the entire picture of the field.

To gain some intuition for this, consider the case $f(z) = V_0 \, \ln (z/R)$, where $V_0, R$ are two constants with dimensions of potential and length, respectively.  In terms of polar coordinates $z = \rho \, e^{i\phi}$, $f(z)=V_0 \, \ln(\rho/R)+i{V_0} \phi$.
The equipotential lines are $\rho=\text{const.}$ and the field lines are $\phi=\text{const.}$ --- i.e.
the field is directed radially. We recognize this field --- it is simply the planar cross section of the field due to an infinite line of charge in three dimensions. This is as expected, because a two-dimensional problem is equivalent to a three-dimensional one with a symmetry in the third direction. In two dimensions, this is the field due to a point charge. The parameter $V_0$ determines the strength of the linear charge density (in three dimensions) or the charge (in two dimensions) and $R$ is a parameter related to the reference point for the potential. The reader is encouraged to find $f(z)$ for some other familiar fields. The essential point is that two-dimensional electrostatics is most naturally formulated in terms of holomorphic quantities.

\section{Fields on the Plane}

A consequence of the existence of $f(z)$ is that electrostatics is conformally invariant. Since the composition of two holomorphic functions is again holomorphic, this means that under a conformal transformation, an electric field gets mapped into an electric field. In fact, the full set of Maxwell equations are conformally invariant, but this is subtler to show. 

To gain familiarity with the complex potential, we will work this out explicitly and relate two different fields by a conformal transformation.

Let us consider the example of a constant electric field $E_0$ pointing in the $+x$ direction in the plane. This could be sourced, for example, by a line of charge parallel to the $y$-axis (or a sheet of charge in three dimensions which is perpendicular to the $xy$-plane and contains the $y$-axis), far away from the region of interest. The complex potential is given by 
\begin{equation*}
    f(z) = -E_0 \, z.
\end{equation*}

Consider now the conformal mapping\footnote{To be precise in terms of dimensional analysis, we should write $w/L=\exp(z/R)$, where $L,R$ are two lengths. Without loss of generality, we can always assume that $L=R=1$ or that the coordinates $z, w$ are really the dimensionless ratios $z/R$, $w/L$.}
 $w = e^z$. Let us consider the image of the field in the $w$-plane. Since $w = e^z = e^{x + i y} = e^x e^{iy}$, we see that the lines $x = C$ are mapped to \textit{circles} of radius $e^C$ in the $w$-plane, and lines of $y = C$ are mapped to \textit{rays} at an angle $C$ with the positive $x$-axis. In particular, the field lines point out radially. The complex potential in the $w$ plane is 
\begin{equation*}
    g(w) = f(z(w)) = -E_0 \,  \ln w. 
\end{equation*}
This is exactly the complex potential for a point charge, if we identify $-E_0 = V_0$\footnote{This equation seems to violate dimensional analysis, but we must recall that we have chosen units by setting a length scale to unity, as discussed in the previous footnote.}. 

Something peculiar has happened: we started with a field sourced by a line of charge in two dimensions and finished with a point charge sitting at the origin. Where did this charge come from? To understand this subtlety, we may consider placing the the line of charge along some line $x = x_0$ in the $z$-plane. That is, we take the line to be at a finite distance, instead of infinitely far away. The image of this in the $w$-plane is a circle of radius $e^{x_0}$. The field in the $w$-plane corresponds precisely to the field due to this circle of charge\footnote{In three dimensions, this would represent the cross-sectional view of a cylindrical configuration.}. Consider now the limit $x_0 \to -\infty$, when the source is infinitely far away in the $z$-plane. In the $w$-plane, this is the zero-radius limit of the circle, which is a point charge.

The ``ghost charge" at the origin in the $w$-plane seems to spontaneously appear, but as we saw from the physics, it is merely a limit.
It is the necessary charge density to source the field under consideration. We can state this mathematically as follows. The exponential map $w = e^z$ does not send the $z$-plane to the $w$-plane surjectively. Its image is $\mathbb{C} \smallsetminus \{0 \}$, the $w$-plane minus the origin. When we formally add back the point in $w = 0$ as a limit point, a ghost charge appears there.

We aim to describe electric fields on more general domains, namely compact surfaces. For compact surfaces, it turns out that ghost charges are \textit{required} by topology --- on a generic surface, one cannot have a sourceless field. It would be beyond our scope to prove this, but the reader may hopefully develop some intuition from our treatment of the examples. 

Let us develop some results which will be useful in later sections. Recall that electromagnetism is linear. We may therefore use the superposition principle, and add solutions together to find other solutions. Given a positive charge at $z=a$ and a negative charge at $z=-a$ of equal strength, the complex potential 
$$
     f(z) = - V_0 \, \ln (z-a) + V_0 \, \ln(z+a) = - V_0 \, \ln{z-a\over z+a} 
$$
describes an  electric dipole created from two charges separated by length $2a$. To find the complex potential of a fundamental dipole, we
take the limit $a\to 0$ at the same time with $V_0\to 0$ such that the product $V_0 \, (2a)=p$ remains finite. At this limit
$$
    f(z) = - {p\over z} .
$$
Similarly, we can take two opposite dipoles at $z=\pm a$:
$$
     f(z) = -p\, \left( {1\over z-a} -{1\over z+a} \right) = {-2a\,p \over (z-a)\,(z+a)}.
$$
In the limit $a\to0$ with $p\to0$ such that $2ap=Q$ remains finite, we find the quadrupole complex potential:
$$
     f(z) = -{Q\over z^2} .
$$ 
The negative sign in the previous potentials is not important; it  relates to the relative orientation of the charges. If we interchange the position of the charges or dipoles, $a\mapsto-a$, then the opposite sign appears.
We may continue in this fashion to find higher-order multipole fields. In particular, it is easy to verify that multipoles come in the form
$$
     f(z) = {M_n\over z^n} ,
$$ 
with $n=1,2,\dots$ and $M_n$ a constant that defines the strength of the $(2n)$-pole. We will call $M_n$ the moment of the  
 $(2n)$-pole; for the dipole, $M_1=p$ is the dipole moment and for the quadrupole, $M_2=Q$ is the quadrupole moment.
We have discovered our key rule: specifying the order of a pole in the complex function $f(z)$ is equivalent to specifying a multipole source for the field. A function which is holomorphic everywhere, with the exception of a finite set of poles, is called \textit{meromorphic}. Liouville's theorem in complex analysis states that a meromorphic function is uniquely fixed by its limiting behavior at its poles and zeroes. Finally, recall that a solution to an electrostatics problem with given boundary conditions is unique. It does not matter how we discover a solution; once we do, we know that it is the only solution. This, together with Liouville's theorem, will allow us to completely determine the fields. 

\subsection*{Multipole expansion}
For the purpose of completeness, and to connect these ideas to the standard topics a student encounters in the electromagnetism texts, we briefly review the notion of multipole expansion and its connection to complex analysis.

Given a general distribution of charge (described by the  charge density $\rho(\vec{r})$), the formal solution for the electrostatic potential in three dimensions 
may be written as an integral 
\begin{equation}
   \varphi(\vec{r}\,) = \frac{1}{4\pi \epsilon_0} \iiint d^3 \vec{r}^{\,\prime} \frac{\rho(\vec r^{\,\prime})}{|\vec{r} - \vec{r}^{\,\prime}|}.
\label{eq:Multipole1} 
\end{equation} 
This merely reflects the superposition principle:  the distribution of charge is a sum of infinitesimal charges parametrized by their location $\vec r^{\,\prime}$, each creating an infinitesimal potential at the point $\vec r$. If one now imagines that the charge density is confined in some finite region of space,
then for  length scales large relative to the size of the distribution\footnote{Actually, this assumption is not necessary. Using standard identities for Legendre polynomials, the integrand can be expanded in a series for any relation of the $r$ and $r'$.}, the ratio $r'/r \ll 1$ (where $r'$ and $r$ denote the magnitudes of $\vec{r}^{\,\prime}$ and $\vec{r}$ respectively).
Expanding the denominator in terms of this ratio, one then finds that the potential has an expansion of the form 
\begin{equation*}
        \varphi(\vec{r}\, ) = \frac{Q}{4\pi \epsilon_0 r} + \frac{P}{4\pi \epsilon_0 r^2} + \cdots . 
\end{equation*}
The coefficients in this expansion are the so-called multipole moments of the distribution: The first term is the potential of the distribution if it was a point charge, the second adds a dipole correction, and so on. 
In this way one sees that, at least on a formal level, it is sufficient to consider multipole fields to solve electrostatics.

In two dimensions, 
given  a distribution of charge with surface density $\sigma(\vec r)$, we can use superposition principle to find the potential:
\begin{equation*}
     \varphi(\vec{r}) = -\frac{1}{2\pi \epsilon_0}   \,    \iint dA' \,  \, \sigma(\vec r^{\,\prime})    \,   \ln|\vec r - \vec r^{\,\prime}| ,
\end{equation*}
Similarly, the complex potential is
\begin{equation}
     f(z) = -\frac{1}{2\pi \epsilon_0}   \,    \iint dA' \,  \, \sigma(z',\bar z')    \,   \ln(z - z') .
\label{eq:Multipole2} 
\end{equation} 
Using the Laurent expansion of the logarithmic function for $|z'/z|<1$, 
$$
         \ln(1-z) = -\sum_{n=1}^\infty {z^n\over n} ,
$$
we easily find
\begin{equation}
     f(z) =    -a_0  \, \ln z  +  \sum_{n=1}^\infty      {a_n \over z^n },
\label{eq:Multipole3} 
\end{equation}
with
$$
       a_0={1\over2\pi \epsilon_0}  \,  \iint dA' \,  \, \sigma(z',\bar z') , \quad   
       a_n =  {1\over2\pi \epsilon_0 n}  \,  \iint dA' \,  \, \sigma(z',\bar z')    \,  z^{\prime n}   .
$$
Equation \eqref{eq:Multipole3} is the multipole expansion of the 2-dimensional distribution. The first term is a point charge located at the origin, while each additional term adds a correction in terms of a multipole. In this way we see a connection between the multipole expansion in two dimensions and the standard Laurent series\footnote{Since equation \eqref{eq:Multipole3} contains a logarithm, strictly speaking, it is not a Laurent series. However, the reader  understands how to convert it to an honest Laurent series with one additional step.}
of a meromorphic function.

\section{Fields on the Sphere}

Now that we have gained some experience with our basic principle --- that conformal maps take fields to fields, and limit points imply ghost charges --- we can explain how to do electrostatics on the sphere. We begin by explaining how to describe the sphere in complex coordinates. The key is a geometric construction known as stereographic projection.

Before continuing, we address a question that the alert reader may have. Electrostatics on the sphere could just as well be studied  by using the ambient three-dimensional space, parameterized in spherical polar coordinates $(r, \theta, \phi)$, and keeping $r$ constant. We would then obtain the potential as a function of the angles $(\theta, \phi)$. Although such an approach is possible, it is  analytically complicated and ignores some fundamental aspects of the topic. The geometrical approach we present is much more efficient. We will still use the ambient space, but in a subtler way --- it will allow us to obtain a more convenient parametric description of the sphere. The reader will understand these kind of subtleties in more detail if he/she studies the general theory of manifolds.

Consider a sphere with radius $R$ in three dimensions, 
$$
         X^2+Y^2+Z^2=R^2.
$$
$X, Y, Z,$ are three-dimensional Cartesian coordinates. We use capital letters to avoid confusion with the complex coordinate $z=X+iY$ on the $XY$-plane. Without loss of generality, we will set $R=1$. The north pole, in Cartesian coordinates $(X, Y, Z)$, is $\text{N} = (0, 0, 1)$. We let $(r, \theta, \phi)$ be the standard spherical coordinates. Hence the points of the sphere are parametrized as $(1,\theta, \phi)$ or simply
$(\theta,\phi)$; two parameters are sufficient since the sphere is a 2-dimensional surface.

Given any point P on the sphere, draw the line NP. Denote the point of intersection of NP with the $XY$-plane as $\text{P}'$. This point $\text{P}'$ may be described in polar coordinates $(\rho, \phi)$ in the $XY$-plane. Taking a cross-section, by elementary trigonometry and similar triangles one deduces that the distance in the $XY$-plane from the origin to $\text{P}'$ is $\rho =  \cot( \theta/2)$. Note the polar angle $\phi$ in the plane is the same as the 3-dimensional azimuthal angle $\phi$. Now, we regard the $XY$-plane as the complex plane, with coordinate $z = \rho e^{i\phi}$, so that we can write the stereographic projection as
\begin{equation*}
       z =  e^{i\phi} \,  \cot\Big( \frac{\theta}{2} \Big).
\end{equation*}

\begin{figure}[h!]
\begin{center}
   \setlength{\unitlength}{1mm}
    \begin{picture}(140,65)
     \put(0,0){\includegraphics[width=14cm]{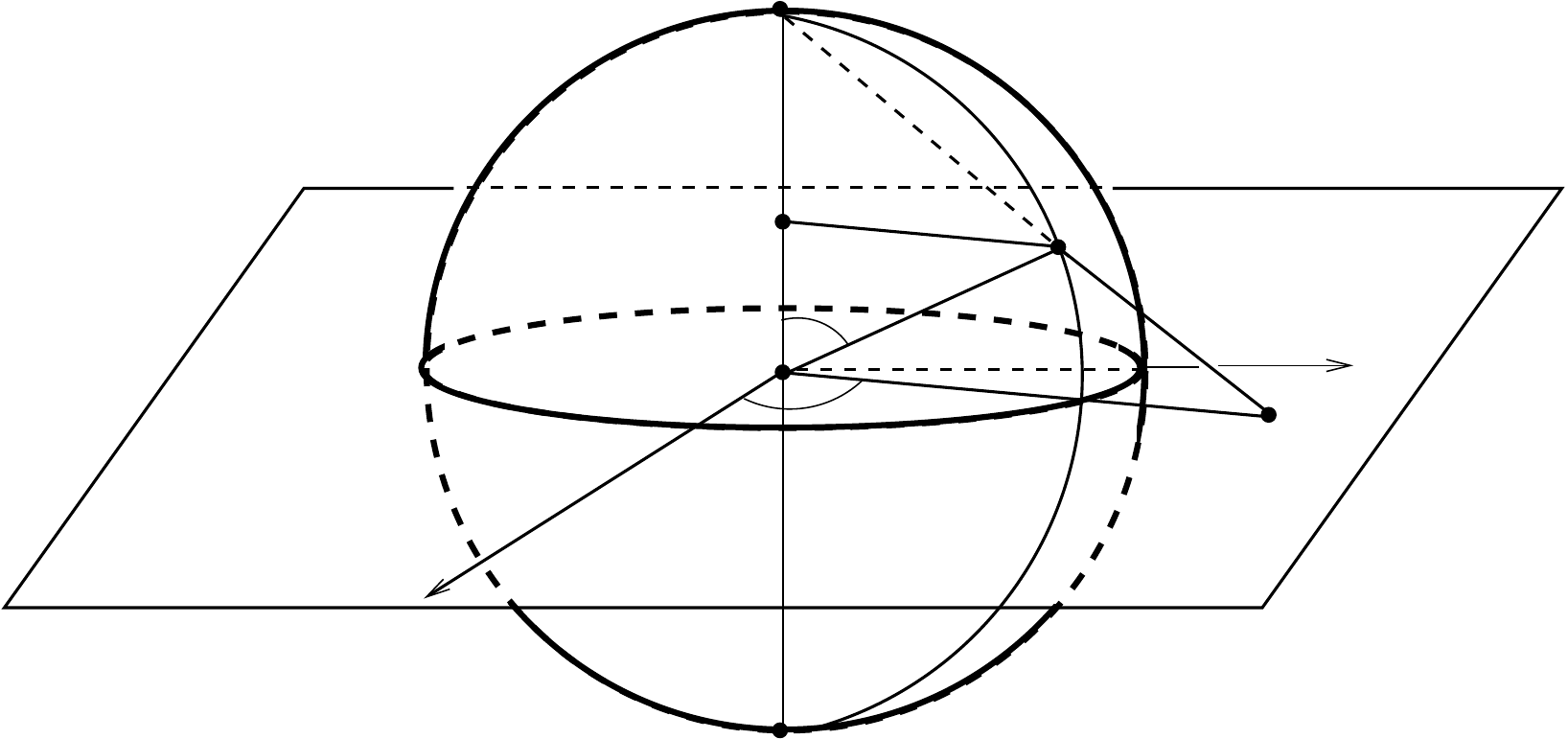}}
     \put(69,66.5){N}
     \put(69,-3){S}
     \put(66,33){O}
     \put(95,45){P}
     \put(115,28){P$'$}
     \put(35,13){$X$}
     \put(118,35){$Y$}
     \put(73,26.5){\colorbox{white}{\small $\phi$}}    
     \put(72.7,38.3){\colorbox{white}{\small $\theta$}}  
\end{picture}
\end{center}     
\caption{Stereographic projection of the sphere.}
\label{fig:StereographicProjection}
\end{figure}
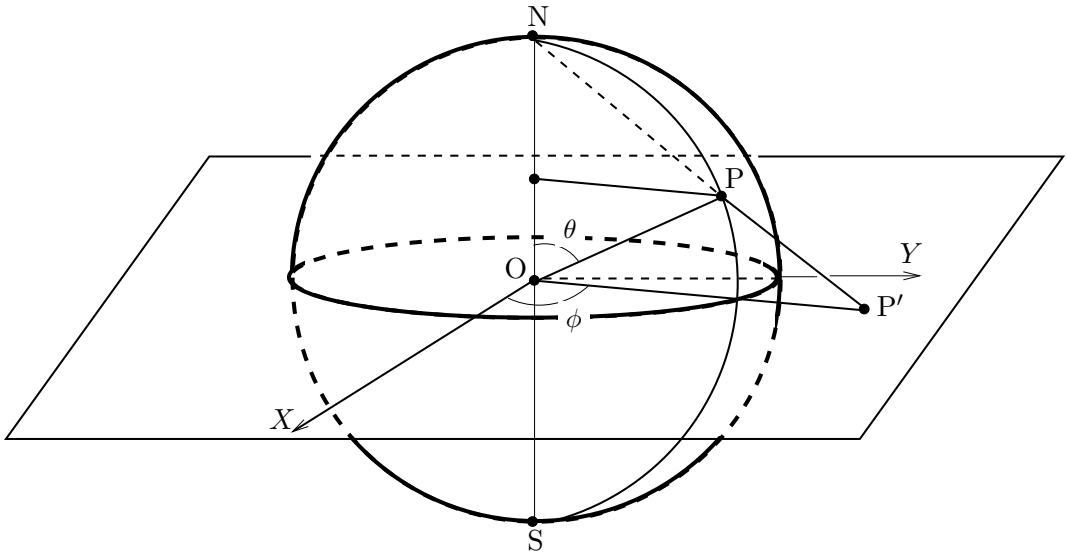

Hence, given the point on the sphere with angular coordinates $(\theta,\phi)$, the corresponding point in the $XY$-plane, regarded as the complex plane, has coordinate $z$ given by the above formula. This equation is valid for every point on the sphere, except for the north pole --- as we approach $\theta = 0$, $|z| \to \infty$. The stereographic projection has `unwrapped' the sphere minus a point to the plane. Turning this around, we may consider the sphere as the plane, together with ``the point at infinity", namely the north pole. This is analogous to what we did in the previous section, but in that case we added the origin as a limit point. Since the sphere is compact, mathematicians say that the sphere gives a \textit{one-point compactification} of the plane.

The theory of manifolds provides a concrete set-up which specifies how to deal with surfaces that have points which escape from our mapping, which in this case is the local coordinate $z$. Without going into the details, when this happens, a second choice of local coordinate is necessary in a neighborhood of such points. If the local coordinates we have introduced do not cover the surface, we add a third map, and so on. For the stereographic projection of the sphere, a second map is all we need. That is, we would like to have another coordinate description of the sphere which includes the north pole. 
This can be accomplished if we  stereographically project from the south pole $\text{S}= (0, 0, -1)$. Once again, given a point P on the sphere, we form the line SP, which intersects the $XY$-plane at some $\text{P}''$. If we use a complex coordinate $w$ in the plane to denote the coordinates $\text{P}''$, supposing P has angular coordinates $ (\phi,\theta)$, we obtain that 
\begin{equation*}
        w = e^{-i\phi} \tan\Big( \frac{\theta}{2} \Big).
\end{equation*}
The sign on $\phi$ must be reversed in order to have a consistent orientation on the sphere.  

Notice that any point P of the sphere which is not the north or the south pole can be described with any of the two complex coordinates $z$ or $w$,
such that $z, w \neq 0$, 
$$
                    w = {1\over z} .
$$
 This means that our two coordinates are in fact holomorphic functions of one another: 
 mathematicians would say that we have given the sphere the structure of a \textit{complex manifold} --- we have covered it with two open neighborhoods, each with a choice of complex coordinate (commonly referred to as the $z$-patch and $w$-patch), such that the two choices of coordinate are related by a holomorphic change of variable. The sphere, when described in terms of complex coordinates, is known as the \textit{Riemann sphere}. It gives us a precise way to talk about infinity in the complex plane --- simply work in the $w$-patch and consider $w = 0$. 

How does this construction relate to physics? The key is that the stereographic projection is in fact a conformal map. To verify this, recall that the infinitesimal distance $ds$ between points on a sphere is given by $ds^2 = d\theta^2 + \sin^2\theta \, d\phi^2$. By relating the differentials $d\theta$ and $d\phi$ to $dz$ and $d\overline{z}$ one obtains, after some algebra: 
\begin{equation*}
     ds^2 = \frac{4}{(1 + |z|^2)^2} dz d\overline{z}.
\end{equation*}
From this expression, we can see that the stereographic projection is conformal --- the infinitesimal distance squared on the plane is simply $dz d\overline{z} = dX^2 + dY^2$, and on the sphere we find the same infinitesimal distances, simply rescaled by a position-dependent scale factor. Since rescaling preserves angles, this local rescaling preserves angles at every point, so the mapping is conformal. 

Conformal invariance then implies that the electrostatic equation takes the remarkably simple form on the Riemann sphere
\begin{equation*}
    \partial_z \partial_{\overline{z}} \varphi = 0.
\end{equation*}
In particular, it is identical to what we found on the complex plane. We may therefore use the complex potential. This means that given any field line pattern on the plane, we simply invert the stereographic projection to obtain the corresponding field line pattern on the sphere. Since stereographic projection is conformal, we know that this procedure will produce a good electric field on the sphere. 

Let us do some examples. First, we take a point charge $+q$ sitting at the south pole, $z = 0$. The complex potential is 
\begin{equation*}
           f(z) = -\frac{q}{2 \pi \epsilon_0} \ln z.
\end{equation*}
We have used a more standard choice of parameters for this physical example--in the notation of the previous sections, $V_0 = -q/2\pi \epsilon_0$. From the stereographic projection, we see that the field lines run along the lines of longitude of the sphere. Since we are on the sphere, we have to check the behavior at infinity, that is, in the $w$-patch near $w = 0$. We see that 
\begin{equation*}
       f(z(w)) = -\frac{q}{2 \pi \epsilon_0} \ln \Big( \frac{1}{w} \Big) = + \frac{q}{2\pi \epsilon_0} \ln w. 
\end{equation*}
We see another point charge, this time of charge $-q$, a sink in the field. This is a ghost charge that has appeared at our limit point $w = 0$. Its appearance is topologically necessary for a non-singular field on the sphere. We conclude that it is impossible to have an isolated point charge on a sphere --- topology demands that we obtain a dipole.  In fact, one of the results of the general theory of Riemann surfaces implies that it is impossible to have an isolated point charge on any surface at all. The charges must always sum to zero. Intuitively this statement is obvious: imagine some positive 
charges on the surface. These are sources of electric field lines. The latter must terminate somewhere on the surface. But Riemann surfaces (like the Riemann sphere) are compact, in particular finite. Therefore, the lines cannot go out to infinity. There must necessarily exist sinks (negative charges) to allow the lines to terminate. We give a proof of this result, in the simple case of the sphere, at the end of the section. 

As a second example, let's take 
the field given by $f(z) = -E_0z$ on the plane, which we considered in the last section
and check its behavior at infinity on the Riemann sphere. We see that 
\begin{equation*}
     f(z(w)) = -\frac{E_0}{w}.
\end{equation*} 
This is the complex potential produced by the dipole. 
Thus, the constant field induces a \textit{ghost dipole} at infinity. Continuing in this fashion, it is easy to see that by considering potentials 
$f(z)\sim z^n$, we end up  with ghost multipole fields.
Hence, the procedure to solve electrostatics on the Riemann sphere is equivalent to specifying the zeroes and poles of the function $f(z)$.

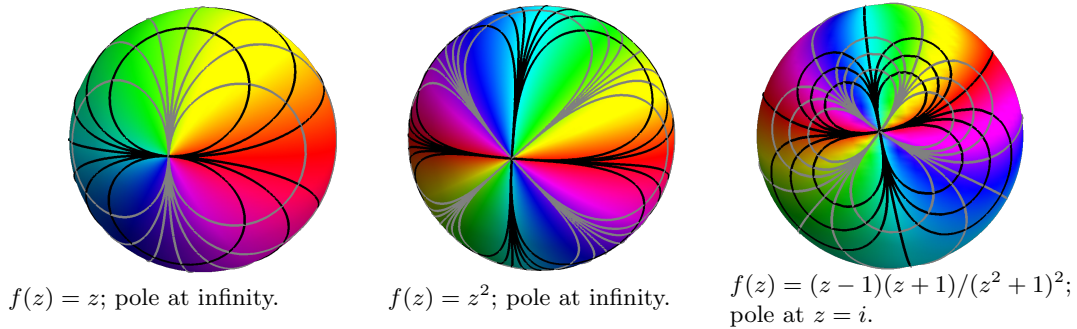
\begin{figure}[h!]
\begin{center}
   \setlength{\unitlength}{1mm}
    \begin{picture}(130,45)
     \put(0,0){\includegraphics[width =4.1cm]{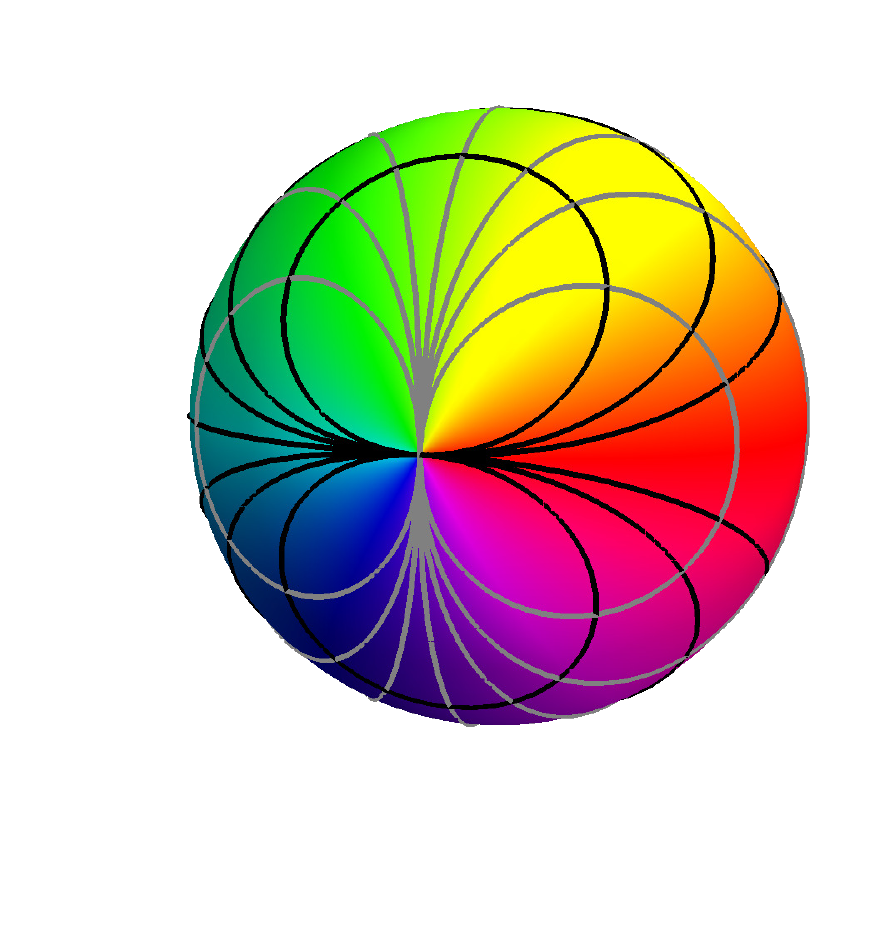}} 
     \put(45,0){\includegraphics[width =4cm]{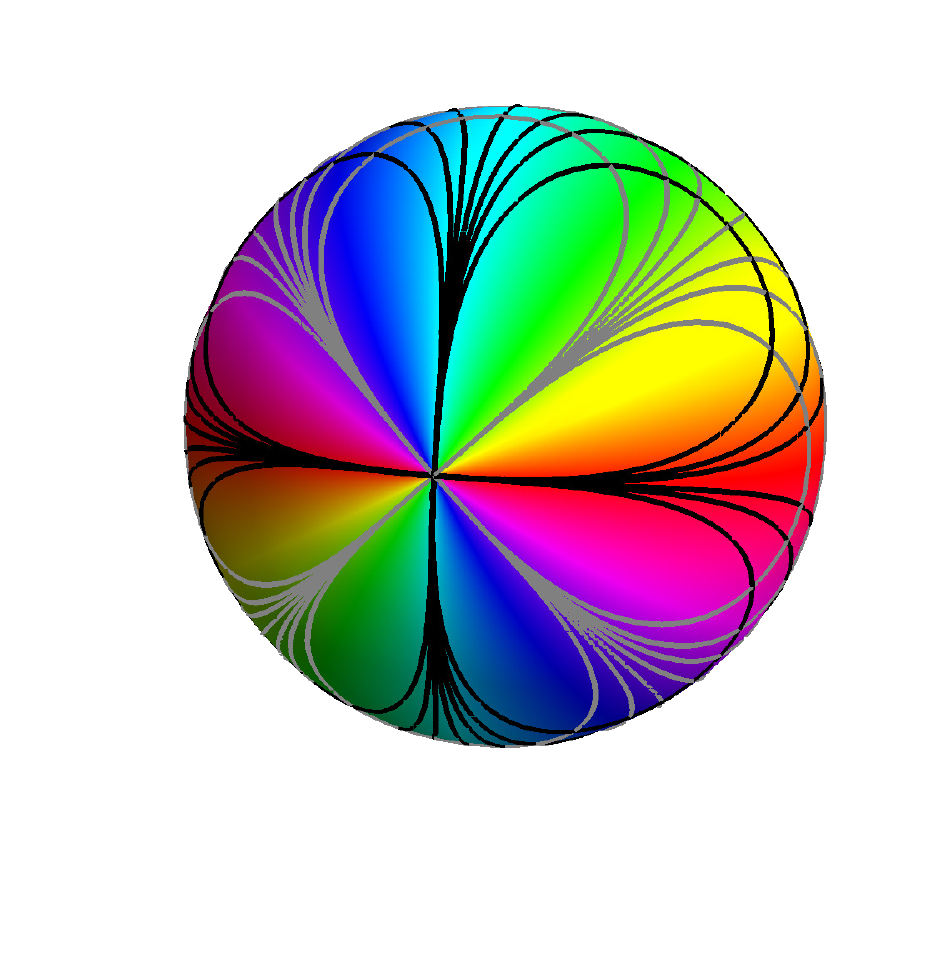}}
      \put(90,0){\includegraphics[width =4.2cm]{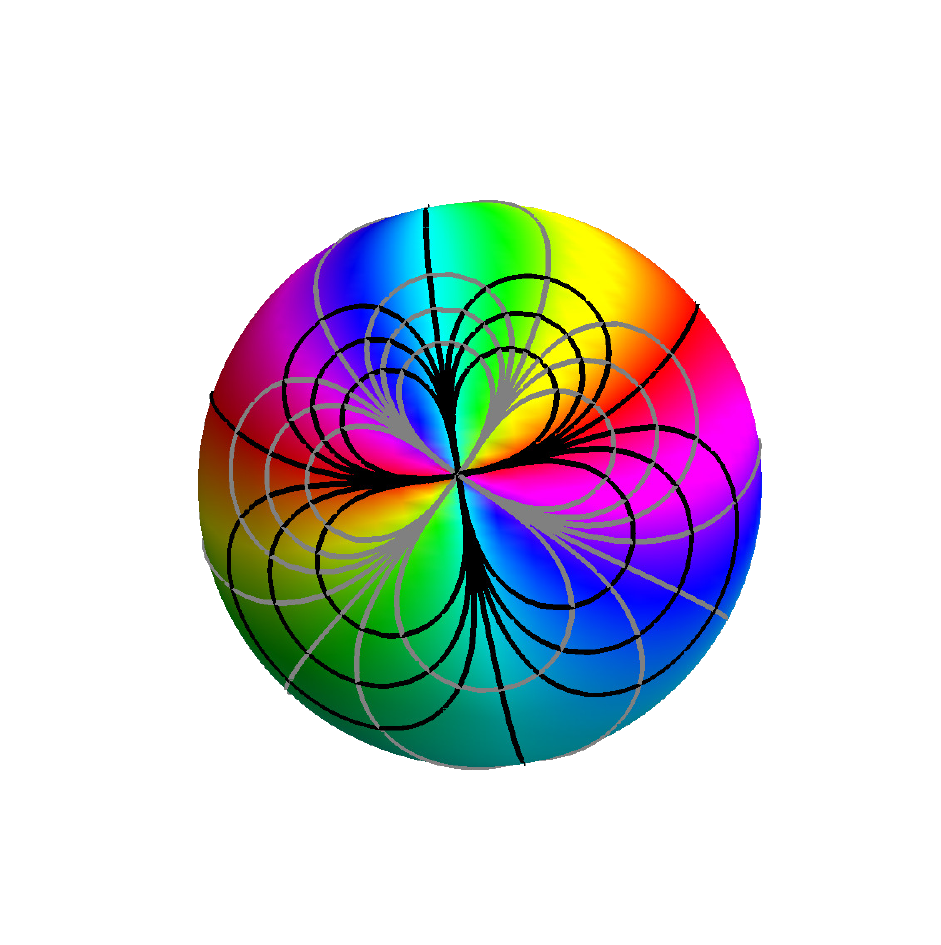}}
     \put(-5,-2){\footnotesize $f(z) = z$; pole at  infinity.}     
     \put(45,-2){\footnotesize $f(z) = z^2$; pole at infinity.}        
     \put(90,-2){\footnotesize \parbox{4.5cm}{$f(z) = (z-1)(z+1)/(z^2 +1)^2$; pole at $z = i$.}}
    \end{picture}
\end{center}
\caption{Various fields on the sphere:  a dipole, quadrupole, and some rational function. Field lines are in black, equipotentials are gray. The hue corresponds to the phase of $f(z)$.}
\label{fig:spherefields}
\end{figure}

In this section, we have sketched a physical explanation of the fact that the poles and zeroes of the function correspond to the sources of the electric field. And we know that, given the charge distribution, the electric field is fixed uniquely. We draw some examples in Figure \ref{fig:spherefields} so that the reader can have a visual representation of the mathematical statements.

\subsection*{Addendum on the ghost dipole}
As somewhat of an aside, we remark that we have so far used the field $f(z) = -E_0 z$ in two ways in this paper: in the preceding section, we explained how it induces a ``ghost charge" under conformal mapping, and in this section, we have shown that it induces a ``ghost dipole" under a different conformal mapping. A natural question is whether it is possible to explain the appearance of this dipole as the limit of a finite-size object in a manner similar to the way that we explained  the appearance of the charge. In fact this is the case, and the explicit construction is as follows. 

Consider ordinary electrostatics in the plane with a circle of radius $R$ of charge density\footnote{The penetrating student will recognize
an underlying dipole immediately: the total charge of the ring is zero while its two hemicircles are charged oppositely. We invite readers, as an exercise, to write down the charge distribution that will give rise to an arbitrary multipole, and work out the analysis in this case.} 
$\lambda(\theta) = - \lambda_0 \cos \theta$. By standard techniques, one solves for the potential as
\begin{align*}
\varphi(\vec{r}\, )  =  \begin{cases}
                                         -\frac{\lambda_0}{2\epsilon_0} \, r \,  \cos \theta ,  & \text{if }r\le R, \\
                                         -\frac{\lambda_0 R^2}{2\epsilon_0} \,  \frac{\cos \theta}{r},   &\text{if }r\ge R.
                                   \end{cases}
\end{align*}
The complex potential is then
$$    
       f(z)  =   \begin{cases}
                                          -\frac{\lambda_0}{2\epsilon_0} \, z,  & \text{if }r\le R, \\
                                          -\frac{\lambda_0R^2}{2\epsilon_0} \, {1\over z},   &\text{if }r\ge R.
                                   \end{cases}
$$ 
In this particular case, the form of the complex potential is obvious given the explicit expression of the potential $\varphi$. It is also possible to calculate it from first principles, using the concepts developed in the previous section. Consider again superposing the potentials of each point of the ring:
\begin{equation*}
     f(z) = -\frac{1}{2\pi \epsilon_0}   \,    \int d\ell' \,  \, \lambda(z',\bar z')    \,   \ln(z - z') ,
\end{equation*}
For $r\gg R$, we Laurent expand the logarithm in terms of $z/z'$. This leads to
\begin{equation*}
     f(z) =    -a_0  \, \ln z  +  \sum_{n=1}^\infty      {a_n \over z^n },
\end{equation*}
with
$$
       a_0={1\over2\pi \epsilon_0}  \,  \int d\ell' \,  \, \lambda(z',\bar z') , \quad   a_n =  {1\over2\pi \epsilon_0 n}  \,  \int d\ell' \,  \, \lambda(z',\bar z')    \,  z^{\prime n}   .
$$
In this case $d\ell'=Rd\theta'$, $z'=R\,e^{i\theta'}$. Then $a_n=0$ for all $n\ne1$ and $a_1=-{\lambda_0 R^2\over 2 \epsilon_0}$. (The integrals, when written in terms of $\theta'$, are basic.). Similarly, we can work out the result for $r\le R$.
Incidentally, notice the continuity of the potential on the ring; the electric field does not have to be continuous on the ring since charge is the source of field lines.

Let the ring grow on the plane, that is, $R \to \infty$ in the $z$-patch.  In this limit, the field for $r \geq R$ becomes inaccessible to an observer near the origin, and we recover the field $f(z) = -E_0z$ considered previously. 

However, in terms of the stereographic projection, as $R \to \infty$ we have a shrinking circle approaching the north pole on the Riemann sphere.  In the $w$-patch, $z\mapsto w=1/z$, the field is 
$$    
       f(w)  =   \begin{cases}
                                          -\frac{\lambda_0 R^2}{2\epsilon_0} \, w,  & \text{if }|w|\le 1/R, \\
                                          -\frac{\lambda_0}{2\epsilon_0} \, {1 \over w},   &\text{if }|w|\ge 1/R.
                                   \end{cases}
$$ 
As $R \to \infty$ the region $|w| \leq 1/R$ disappears and one is left simply with the dipole field $-E_0/w$ at infinity, as before. We see here that the dipole at infinity has arisen as the limit of a shrinking circle.

\subsection*{Zero total charge on sphere}
For completeness, we provide a proof that it is impossible to have an isolated point charge on the surface of the sphere. We have chosen to confine ourselves to the sphere in order to avoid using more mathematical results than necessary. The interested reader can easily consult the references to fill in the mathematical details for  a general Riemann
surface.

Consider a configuration with $n$ point charges of charge $q_i$ located at points $z_i$ on the surface of the sphere. 
 This complex potential of the configuration is
\begin{equation*}
    f(z)  =   -\sum_{i=1}^n \frac{q_i}{2\pi \epsilon_0} \ln(z - z_i) . 
\end{equation*}
Consider now the integral of $f'(z)$ on a closed contour $C$ on the sphere which bounds a region enclosing no simple poles of $f'(z)$. Then one has 
\begin{equation*}
    0 = \oint_C f'(z) \frac{dz}{2\pi i},
\end{equation*}
or
\begin{equation*}
    0 =   \frac{1}{2\pi \epsilon_0}   \sum_{i=1}^n  \oint_C   {q_i\over z-z_i} \, \frac{dz}{2\pi i} .
\end{equation*}
For the $i$-th integral, the contour $C$ can be deformed independently to a contour $C_i$ around the pole $z_i$ without including any of the remaining poles:
\begin{equation*}
    0 =   \frac{1}{2\pi \epsilon_0}   \sum_{i=1}^n  \oint_{C_i}   {q_i\over z-z_i} \, \frac{dz}{2\pi i} .
\end{equation*}
Then it is immediate
\begin{equation*}
    0 =  \frac{1}{2\pi \epsilon_0} \sum_{i=1}^n q_i .
\end{equation*} 
So we derive the constraint $\sum_i q_i = 0$ which is impossible to satisfy with a single charge.

\section{Fields on the Torus}

The study of fields on the sphere is interesting geometrically because we may visualize some interesting field line patterns via stereographic projection. However, mathematically it is not so interesting because the complex potentials are the same as those of the familiar fields on the plane. In this section, we explain a beautiful result: the relation of electrostatics to a class of functions one typically encounters when studying the period of a pendulum, for example (or a bead on a hoop--see \cite{BB}). In particular, we embark on  the study of electric fields on the torus (that is, the surface of a donut) which necessarily leads to elliptic functions.

The basic analytic difficulty that we face is to obtain a useful parameterization of the torus. The torus can be characterized as the surface of revolution obtained from taking a round circle $S^1$ in, say, the $XZ$-plane and rotating it about the $Z$-axis. Suppose the circle's center lies on the $XZ$-plane at a distance $a$ from the origin along the $X$-axis, and that the circle has radius $b$, where $0 < b < a$. Then the equation of the torus is 
$$
            \big(a-\sqrt{X^2 + Y^2}\big)^2 + Z^2 = b^2 .
$$ 
If we were to attempt to use vector calculus to study fields on this surface, it would be analytically excruciating. In accordance with our theme, the strategy will be to study it as a 2-dimensional problem and rephrase it in terms of a holomorphic function.

The first step is to search for a conformal map. We give here a geometric construction of the map known as the Clifford embedding. It should be clear that the torus, as a set, is the Cartesian product $S^1 \times S^1$. The embedding of the torus into 3-dimensional space breaks the symmetry between the two $S^1$'s: because of their placement, effectively, we only rotate one of them about the $Z$-axis. The other is unaffected by the rotation. This asymmetry is reflected in the complicated nature of the algebraic equation for the torus in three dimensions. 

The key idea of the Clifford embedding is to find a representation of the torus that preserves the symmetry between the $S^1$'s. The standard way to represent $S^1$ as a round circle of radius $R$ is 
\begin{equation*}
     S^1 = \{ (R\cos \phi, R\sin \phi) \in \mathbb{R}^2 ~|~ \phi \in [0, 2\pi) \} .
\end{equation*}
This naturally lives in a plane, $\mathbb{R}^2$. For later convenience, we will set $R=1/\sqrt{2}$.
So, using the above representation, the natural way to represent the Cartesian product is\footnote{Note that the angular coordinates $(\theta, \phi)$ of this section are completely distinct from the spherical coordinates of the same name in the previous section.} 
\begin{equation*}
     S^1 \times S^1 = \left\{ \left(\frac{1}{\sqrt{2}}\cos \phi, \frac{1}{\sqrt{2}}\sin \phi, \frac{1}{\sqrt{2}} \cos \theta, \frac{1}{\sqrt{2}} \sin \theta\right) 
                                  \in \mathbb{R}^4  ~|~ (\theta, \phi) \in [0, 2\pi) \times [0, 2\pi) \right\}.
\end{equation*}
This torus sits in 4-dimensional space $\mathbb{R}^4$, the set of all $4$-tuples of real numbers $(x_1, x_2, x_3, x_4)$. It is clear that since we have $(x_1, x_2, x_3, x_4) =  ( \cos \phi, \sin \phi, \cos \theta, \sin \theta)/\sqrt{2}$, then from trigonometric identities 
\begin{equation*}
     x_1^2 + x_2^2 + x_3^2 + x_4^2 = 1.
\end{equation*}
This means that the torus $S^1 \times S^1$ is contained within the 3-sphere $S^3$ in four dimensions:
$$
         S^3 = \{ (x_1, x_2, x_3, x_4) \in \mathbb{R}^4 ~|~ x_1^2 + x_2^2 + x_3^2 + x_4^2 = 1\} .
$$
 That is, $S^1 \times S^1 \subset S^3$. This embedding of the torus into $S^3$ is known as the Clifford embedding, and the torus as the Clifford torus.

We now must relate this abstract torus in $S^3$ to the familiar one in $\mathbb{R}^3$. The key, once again, is stereographic projection. We have seen that we may stereographically project an $S^2$ to a plane. This projection can be extended to any sphere, in any number of dimensions. In particular, the 3-dimensional sphere  $S^3$, although harder to visualize,  can be projected stereographically from its north or south pole onto its 3-dimensional equatorial `plane'. 

Let's construct this projection explicitly. Since visualization and Euclidean geometry cannot aid us anymore, we will use a different strategy. First notice that, given that the round $S^3$ has been placed symmetrically inside $\mathbb{R}^4$, the equational `plane' is all points of $\mathbb{R}^4$ with
with $x_4=0$. This describes an $\mathbb{R}^3$. We start with the north pole $\text{N} = (0, 0, 0, 1)$ and consider some point $\text{P} = (x_1, x_2, x_3, x_4) \in S^3$. We consider the line which passes through N and P, and intersects the hyperplane $x_4 = 0$ at some point $\text{P}' = (X, Y, Z, 0)$. We may solve for $X, Y$, and $Z$ in terms of $x_1, x_2, x_3$, and $x_4$ as follows. The line NP is given in parametric form as the set of points
\begin{equation*}
       \text{NP} = \{ (tx_1, tx_2, tx_3, 1 + t(x_4 -1)) \in \mathbb{R}^4 ~|~ t \in (-\infty, \infty) \}.
\end{equation*}
The line $L$ intersects the desired hyperplane when its fourth coordinate vanishes. This happens at $1 + t(x_4 - 1) = 0$, or $t = 1/(1-x_4)$. Inserting this value of $t$ for the other coordinates, we may solve for $(X, Y, Z, 0)$: 
\begin{eqnarray*}
   X & = & \frac{x_1}{1 - x_4}, \\
   Y & = & \frac{x_2}{1 - x_4}, \\ 
   Z & = & \frac{x_3}{1 - x_4}.
\end{eqnarray*}

This is the stereographic projection of $S^3$ minus the north pole down to $\mathbb{R}^3$. We are almost done: if we consider the Clifford torus in $S^3$, $(x_1, x_2, x_3, x_4) = (\cos \phi, \sin \phi, \cos \theta, \sin \theta)/\sqrt{2}$, substituting these values into the stereographic projection gives the following parameterized set in $\mathbb{R}^3$: 
\begin{eqnarray*}
X & = & \frac{\cos \phi}{\sqrt{2} - \sin \theta}, \\
Y & = & \frac{\sin \phi}{\sqrt{2} - \sin \theta}, \\
Z & = & \frac{\cos \theta}{\sqrt{2} - \sin \theta}.
\end{eqnarray*}
This is a somewhat strange parameterization of the set, but some experimentation with the equations reveals that 
\begin{equation*}
     (\sqrt{2} - \sqrt{X^2 + Y^2} )^2 + Z^2 = 1.
\end{equation*}
This identifies this set as the familiar torus in $\mathbb{R}^3$. 

The 3-dimensional version of stereographic projection is also a conformal map. We omit the proof here since the details are unimportant for our purposes.
In four dimensions, the infinitesimal line element is 
$$
    ds^2 = dx_1^2 + dx_2^2 + dx_3^2 + dx_4^2.
$$ 
If we substitute in $(x_1, x_2, x_3, x_4)$ in terms of $\theta$ and $\phi$ on the Clifford torus, we obtain the infinitesimal distance on the Clifford torus: 
\begin{equation*}
     ds^2 = \frac{1}{2}(d\theta^2 + d\phi^2).
\end{equation*}
Remarkably, this is, up to a constant, simply the usual distance on the plane with coordinates $(\theta, \phi)$. Since the angular variables only vary between $0$ and $2\pi$, we have found that the the torus, given by its complicated algebraic equation, conformally maps (via stereographic projection) onto the interior of a square!

This is quite an enchanting mathematical story, but we must return to the physics. At this point, the strategy is familiar. Since the electrostatic equation 
$\laplacian \varphi = 0$ is conformally invariant, we can take electrostatics on the torus and map it to the $(\theta, \phi)$-plane. Introducing the complex coordinate $z = (\phi + i \theta)/2\pi$ on the plane, the equation becomes $\partial_z\partial_{\overline{z}} \varphi = 0$, which can be solved in terms of a complex potential $f(z)$ by $\varphi(z, \overline{z}) = \Re f(z)$. 

The novel feature here is that $\theta$ and $\phi$ are angular variables, so $\theta$ and $\theta + 2\pi$ are considered to be equivalent points, and similarly for $\phi$. In terms of $z$, this means that $z$, $z + 1$, and $z + i$ must be considered equivalent. Thus, for $f$ to be a well defined function on the torus, it must satisfy 
\begin{equation*}
    f(z) = f(z + 1) = f(z + i).
\end{equation*}
In other words, it must be \textit{doubly periodic}. A meromorphic, doubly periodic function is called an \textit{elliptic function}. Clearly, such a function is fixed by its behavior in the ``fundamental square" $(\theta, \phi) \in [0, 2\pi) \times [0, 2\pi)$. 

Actually, the double periodicity may be relaxed slightly: if $f(z + 1)$ and $f(z + i)$ only differ from $f(z)$ by constants, this is still permissible, since it is only the \textit{field} that is physically observable, not the potential. We refer to such functions as quasi-periodic. 

At this point, we may borrow some results from the general theory of elliptic functions to aid in our search for electric fields. We will use the so-called Weierstrass elliptic functions, because their relationship to the physics is more transparent in this case. The basic building block for all elliptic functions is the ``Weierstrass $\wp$-function" 
$$
    \wp(z) = \frac{1}{z^2} + \sideset{}{^*}\sum_{ (m, n) \in \mathbb{Z} \times \mathbb{Z}}   
                     \Big[ \frac{1}{(z + m + in)^2} - \frac{1}{(m + in)^2} \Big],
 $$ 
 where the star in the summation implies that the value $n=m=0$ must be omitted.
 The double sum may be shown to converge. It is doubly periodic by construction --- a periodic shift simply permutes the terms. $\wp(z)$ has a double pole at zero and its periodic equivalents, but is regular everywhere else. It also has has two zeroes in the ``fundamental square", and, of course, their periodic equivalents. 

General results in elliptic function theory ensure that any elliptic function is completely specified by its poles and zeroes. This, in turn, implies that they are all rational expressions in $\wp$ and $\wp'$. The idea is that one may use the poles and zeroes of $\wp$ and $\wp'$ to engineer the divergence or vanishing of the appropriate order at the appropriate points.

Let us finally turn to the computation of some fields. We start with a point charge. The point charge field is characterized by a logarithmic singularity. Since $\wp$ diverges as $z^{-2}$, a point charge $q$ placed at $z = 0$ generates the complex potential 
\begin{equation*}
    f(z) = +\frac{q}{4 \pi \epsilon_0} \ln \wp(z).
\end{equation*}
Since we are on a compact surface, we should not expect the point charge to come alone. In this case, the other sources come from the zeroes of $\wp(z)$. In particular, this complex potential describes a point charge $+q$ at $z = 0$, and two other point charges, each of charge $-q/2$ at the zeroes of $\wp(z)$. 

While it is not possible to have an isolated point charge on the torus, it is in fact possible to have an isolated dipole. If the dipole has dipole moment $p$, the complex potential is 
\begin{equation*}
     f(z) = p \int^z \wp(z') dz' := -p\zeta_W(z),
\end{equation*}
 the antiderivative of $\wp$, often called the Weierstrass zeta function $\zeta_W(z)$. This antiderivative is in fact quasi-periodic, but as we have explained this does not impact the physical field. The way to understand this result is that $\wp(z)$ has a double pole at $z = 0$, but is regular everywhere else, so its integral will have only a simple pole at $z = 0$ (the hallmark of a dipole), and be regular everywhere else. This regularity means that the dipole is in fact isolated. $f(z)$ of the form $\wp'/\wp$ corresponds to a non-isolated dipole.

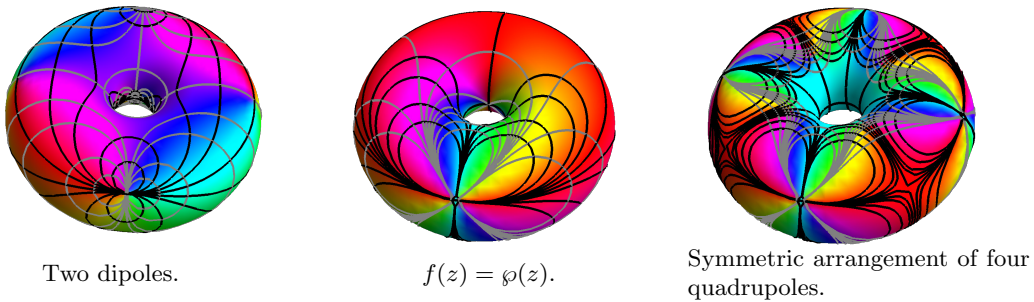
\begin{figure}[h!]
\begin{center}
   \setlength{\unitlength}{1mm}
    \begin{picture}(130,40)
     \put(-8,-2){\includegraphics[width =4.8cm]{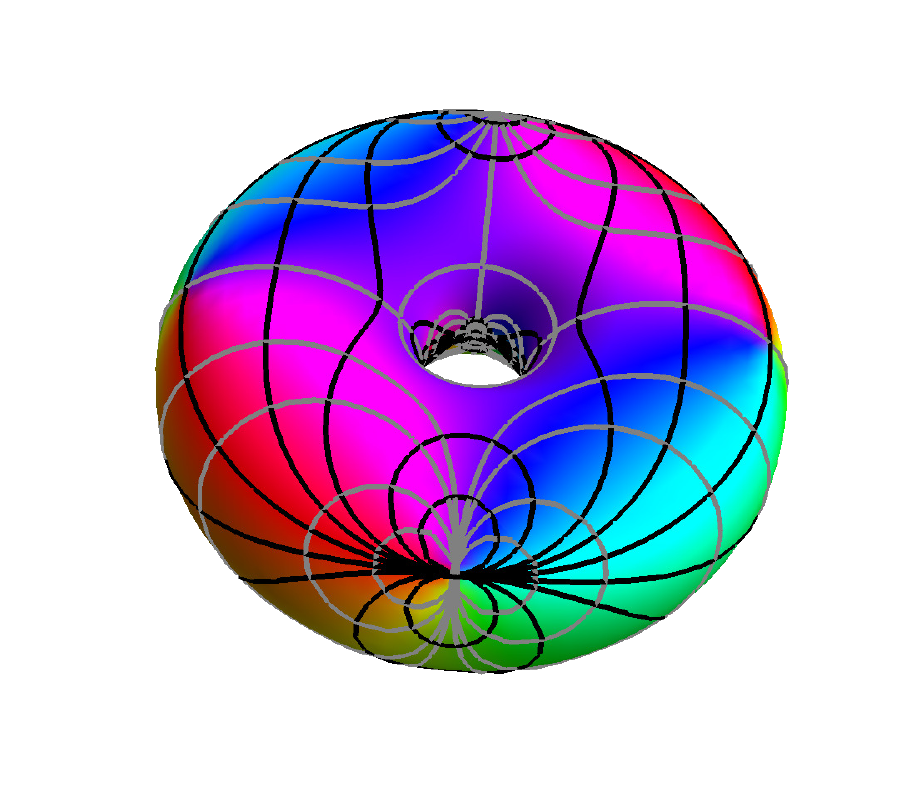}} 
     \put(43,0){\includegraphics[width =4cm]{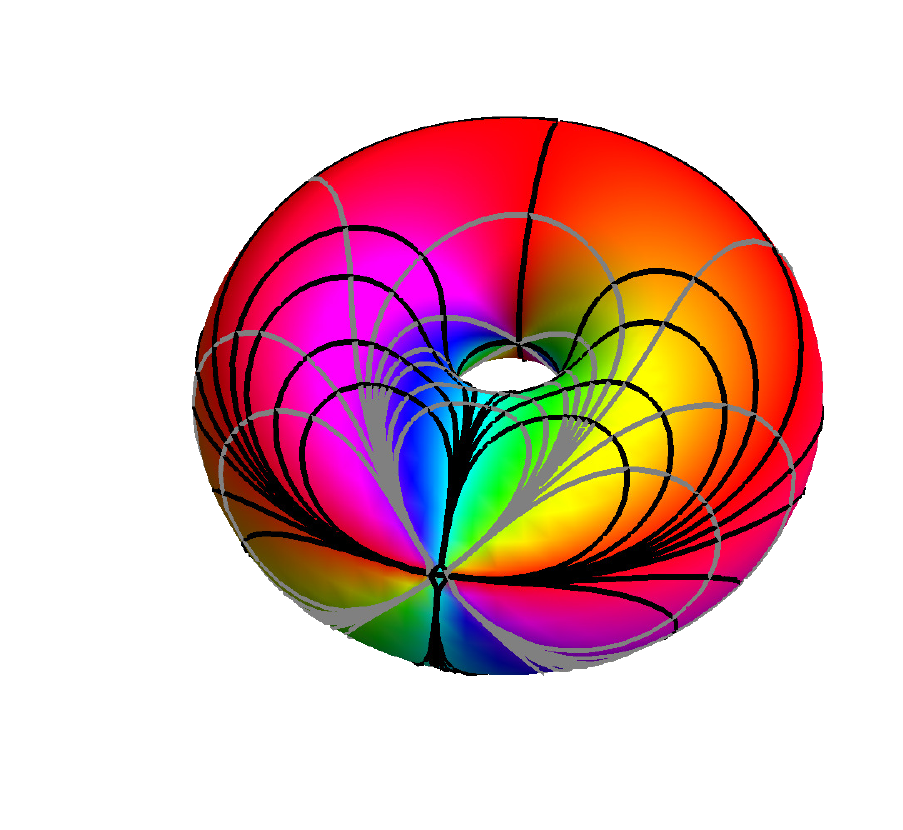}}
      \put(90,0){\includegraphics[width =4cm]{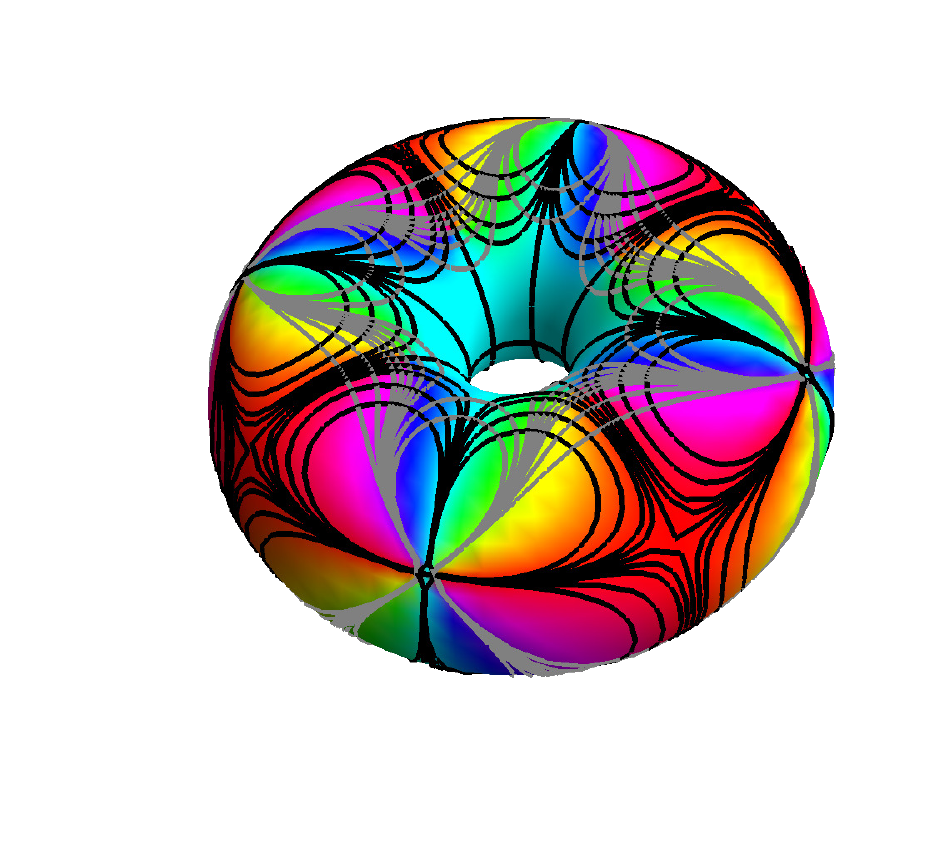}}
     \put(5,-2){\footnotesize {Two dipoles.}}     
     \put(55,-2){\footnotesize {$f(z) = \wp(z)$.}}        
     \put(90,-2){\footnotesize \parbox{4.5cm}{Symmetric arrangement of four quadrupoles.}}
    \end{picture}
\end{center}
\caption{Various fields on the torus--two dipoles, a quadrupole, and four quadrupoles distributed symmetrically. Field lines are in black, equipotentials are gray. The hue corresponds to the phase of $f(z)$.}
\label{fig:OnTorus}
\end{figure}

These fields have been constructed on the abstract Clifford torus, using conformal invariance. Using stereographic projection, one can project them back to the physical torus in $\mathbb{R}^3$, which gives a nice way to visualize the elliptic functions, as we see in Figure~\ref{fig:OnTorus}.
%

\section{Comments \& Conclusion}
We conclude this article by directing the interested reader to the relevant mathematical literature, and pointing out some natural extensions of our constructions in these contexts. 

The main theme, of course, is the mathematics of Riemann surfaces. For a rather accessible introduction that covers an impressive amount of ground, see \cite{eynard}. Classic references on the theory of theta functions on Riemann surfaces (which is really what underlies elliptic function theory and its generalizations) are \cite{fay}, \cite{mumford}. There is also the textbook \cite{farkaskra}, and more mathematically inclined readers may wish to investigate \cite{griffithsharris}. Finally, for the relevant material on electrostatics, see the standard texts \cite{griffiths}, \cite{jackson}. 

For readers that plan on consulting references such as \cite{griffithsharris}, we will explain how our constructions fit into the broader theory. A major theme for us was the fact that a meromorphic function on the surface of interest is essentially characterized by its poles and zeroes --- physically, these corresponded to multipole-type sources for electric fields. One of the central questions in the theory of Riemann surfaces is to what extent this holds in general --- that is, given zeroes and poles on a general Riemann surface, are there any functions with this behavior, and if so, how many linearly independent ones? The answer is provided by the so-called Riemann-Roch theorem and is the natural starting point for the theory of divisors and line bundles in algebraic geometry. 

A related question is, after we have determined that functions exist with the prescribed zeroes and poles, can we construct them explicitly? It turns out that this is the case if one introduces the so-called theta functions. These are naturally defined on an auxiliary space called the Jacobian variety, and the theory of the Abel-Jacobi map explains under what conditions one can ``pull-back" the theta functions to the Riemann surface itself to give concrete descriptions of meromorphic functions. In fact, $\wp(z)$ may be written in terms of theta functions.

We now remark on the natural extensions of our work. Perhaps the most obvious one is an extension of these constructions to Riemann surfaces of higher genus. The inherent difficulty in dealing with such surfaces is the lack of explicit parametric descriptions --- one typically must invoke the full machinery of algebraic geometry and regard the surfaces as projective varieties to make any progress. Nonetheless, it would be interesting to attempt to describe the geometry of higher genus Riemann surface in an approach similar to ours: finding a convenient conformal map which allows for visualization in $\mathbb{R}^3$, giving a concrete representation of the geometric structures on the Riemann surface in terms of electric fields. 

Another possibility is to study the effect of varying the complex structure. Let us briefly explain what this means. Typically, given a real surface, there are actually several inequivalent ways to combine its coordinates into a local complex coordinate $z$. Each way of doing this is called a complex structure, and typically complex structures come in continuous families (so-called moduli spaces). In this language, our Clifford embedding only covered one point in the moduli space, and by introducing additional parameters, it can be generalized to describe inequivalent complex structures. The fields, depending on holomorphic quantities, would correspondingly distort as the complex structure is varied, and this would be an interesting effect. The extension of this to higher genus would be highly nontrivial.

\section*{Acknowledgements}
S.T. thanks C.E. for the opportunity to work on this project, useful discussions, and encouragement.



\begin{thebibliography}{13}
\bibitem{BB}
T. E. Baker and A. Bill, ``Jacobi elliptic functions and the complete solution to the bead on the hoop problem", Am. J. Phys. \textbf{80}, 506-514 (2012). 	

\bibitem{burton}
H.G.A. Burton, A. J. W. Thom, and P-F. Loos, ``Complex adiabatic connection: A hidden non-Hermitian path from ground to excited states", J. Chem. Phys. \textbf{150}, 041103 (2019).
	
\bibitem{EsparzaEtAl}
C. Esparza, P. L. Rend\'on, E. L. Koo,  ``A complete set of two-dimensional harmonic vortices on a spherical surface",
Eur. J. Phys. \textbf{39} (2018) 025709.
	
	
\bibitem{eynard}
B. Eynard, \textit{Lecture Notes on Compact Riemann Surfaces}, \url{https://arxiv.org/abs/1805.06405} (2018).

\bibitem{farkaskra}
H.M. Farkas and I. Kra, \textit{Riemann Surfaces}, 2nd ed. (Graduate Texts in Mathematics, Springer, 1992). 

\bibitem{fay}
J. D. Fay, \textit{Theta Functions on Riemann Surfaces}. (Lecture Notes in Mathematics, Springer Berlin Heidelberg, 1973). 

\bibitem{griffiths}
D. J. Griffiths, \textit{Introduction to Electrodynamics}, 3rd ed. (Prentice Hall, Upper Saddle River, NJ, 1999).

\bibitem{griffithsharris}
P. Griffiths and J. Harris, \textit{Principles of Algebraic Geometry}. (John Wiley and Sons Inc., New York, 1994). 


\bibitem{jackson}
J. D. Jackson, \textit{Classical Electrodynamics}, 3rd ed. (John Wiley and Sons Inc., New York, 1998).

\bibitem{levin}
M. Levin and S. G. Johnson, ``Is the electrostatic force between a point charge and a neutral metallic object always attractive?", Am. J. Phys. \textbf{79}, 843-849 (2011).

\bibitem{MittagStephen}
L. Mittag, and M. J. Stephen ``Conformal transformations and the application of complex variables in mechanics and quantum mechanics", 
Am. J. Phys. \textbf{60} (1992) 207.


\bibitem{mumford}
D. Mumford, \textit{Tata Lectures on Theta Functions}. (Modern Birkh\"{a}user Classics, Birkh\"{a}user, Boston, 1984).	

\bibitem{sokolovski}
D. Sokolovski, S. K. Sen, V. Aquilanti, S. Cavalli, and D. De Fazio, 
``Interacting resonances in the $\mathrm{F} + \mathrm{H}_2$ reaction revisited: Complex terms, Riemann surfaces, and angular distributions", 
J. Chem. Phys. \textbf{126}, 084305 (2007).

\bibitem{staunton}
L. P. Staunton, ``The restoring force on a dielectric in a parallel plate capacitor", Am. J. Phys. \textbf{82}, 853-859 (2014).

\bibitem{weigel}
C. Weigel, J. M Wachter, P. Wagoner, and T. J.
Atherton, “Predicting the influence of plate geometry on the eddy-current
pendulum”, Am. J. Phys. \textbf{84}, 653–663 (2016).
	
\end{thebibliography}
\end{document}